\documentclass[prd,aps,preprint,amsmath,amssymb,showpacs,nofootinbib,superscriptaddress]{revtex4-2}

\usepackage{amsmath}
\usepackage{graphicx,enumerate,xcolor,enumitem,xspace,braket,mathrsfs} 
\usepackage[colorlinks,citecolor=blue]{hyperref}
\usepackage{comment}
\usepackage{float}
\usepackage{booktabs}
\usepackage{cleveref} 
\usepackage{braket}
\allowdisplaybreaks[4]

\newcommand{\tr}{\operatorname{Tr}}

\newcommand{\tev}{\operatorname{TeV}}

\newcommand{\cw}{c_{\widetilde{W}}}
\newcommand{\cwww}{c_{\widetilde{W}WW}}

\begin{document}
\preprint{PITT-PACC-2501}

\title{Maximizing the Azimuthal-Angle Correlation\\  in the Decay of Vector Boson Pairs}
\author{Kun Cheng}
\email{kun.cheng@pitt.edu}
\affiliation{PITT PACC, Department of Physics and Astronomy,\\ University of Pittsburgh, 3941 O’Hara St., Pittsburgh, PA 15260, USA}

\author{Yi-Jing Fang}
\email{fangyj@ihep.ac.cn}
\affiliation{Institute of High Energy Physics, Chinese Academy of Sciences, Beijing 100049, China}
\affiliation{School of Physics, University of Chinese Academy of Sciences, Beijing 10049, China}

\author{Tao Han}
\email{than@pitt.edu}
\affiliation{PITT PACC, Department of Physics and Astronomy,\\ University of Pittsburgh, 3941 O’Hara St., Pittsburgh, PA 15260, USA}

\author{Matthew Low}
\email{mal431@pitt.edu}
\affiliation{PITT PACC, Department of Physics and Astronomy,\\ University of Pittsburgh, 3941 O’Hara St., Pittsburgh, PA 15260, USA}

\date{\today}

\begin{abstract}
Spin correlation observables of particles produced at colliders generally depend on the coordinate basis.  While spin correlations and their associated basis dependence have been studied in top-quark pair production, the basis dependence of spin correlations of $W^\pm$ pairs has not been studied.
In this work, we focus on the azimuthal-angle correlation in the decay of $W^\pm$ pairs, which encodes spin information in the direction transverse to the rotational axis.  We quantitatively describe the basis dependence of the relative azimuthal angle. We further present the optimal coordinate direction to define the azimuthal angle between the decay products of $W^\pm$ bosons to achieve the maximal spin correlation based on symmetry considerations. The results can be used to optimize the measurement of the CP-even and CP-odd interactions, and have applications to quantum information observables in the vector boson pair system. 
\end{abstract}

\maketitle
\newpage

\section{Introduction}
\label{sec:introduction}

One of the most important phenomena in the production of particle pairs at colliders is the correlation between particle spins, which depends sensitively on the fundamental interactions and is often used to probe new physics effects.  In collider environments, massive particles with spin, such as top quarks and $W$ and $Z$ bosons, decay promptly before reaching the detector.  The spin information of these particles is not measured directly but instead propagates to the distribution of their decay products.  Consequently, the spin correlation between two particles is measured from the angular correlation between their decay products.

An important feature of spin correlations is that the observables used to measure them are basis dependent, which has been extensively studied in top-pair production~\cite{Parke:1996pr,Mahlon:1997uc,Uwer:2004vp,Cheng:2023qmz,Cheng:2024btk}.  However, the same basis dependence for the spin correlation of a pair of $W^\pm$ bosons has not yet been studied because its higher-dimensional spin space leads to more complicated spin correlations.  Most generally, the spin correlation matrix of a $W^\pm$ pair has 64 independent entries, compared to the 9 independent entries of the top pair spin correlation matrix.  Together with the net polarizations of the $W^+$ and the $W^-$, a complete measurement of the spin state of the $W^\pm$ pair is a highly demanding 80-parameter fit~\cite{Hagiwara:1986vm}.    

\begin{figure}
  \centering
  \includegraphics[width=0.4\linewidth]{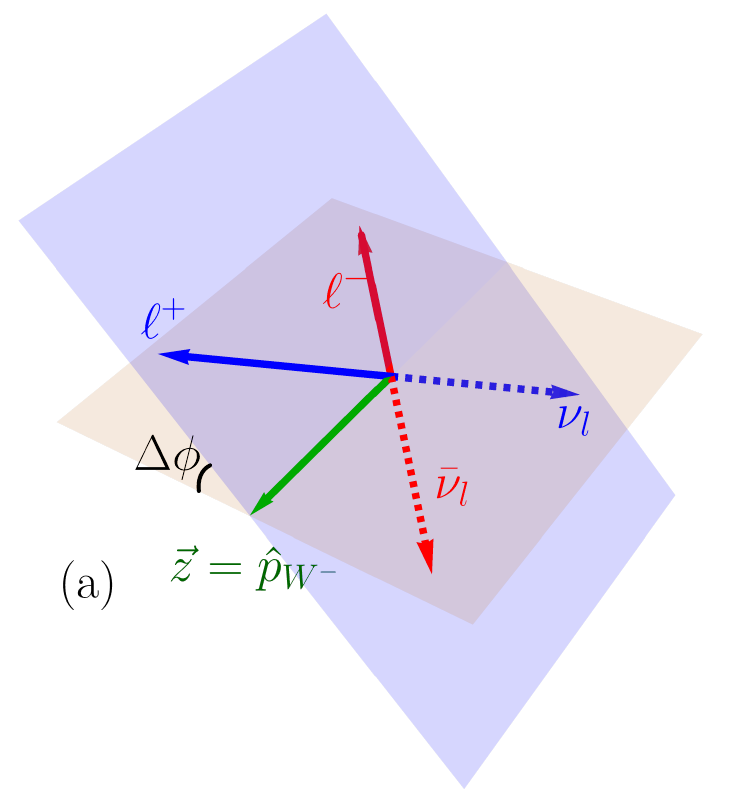}
  \qquad\qquad
  \includegraphics[width=0.4\linewidth]{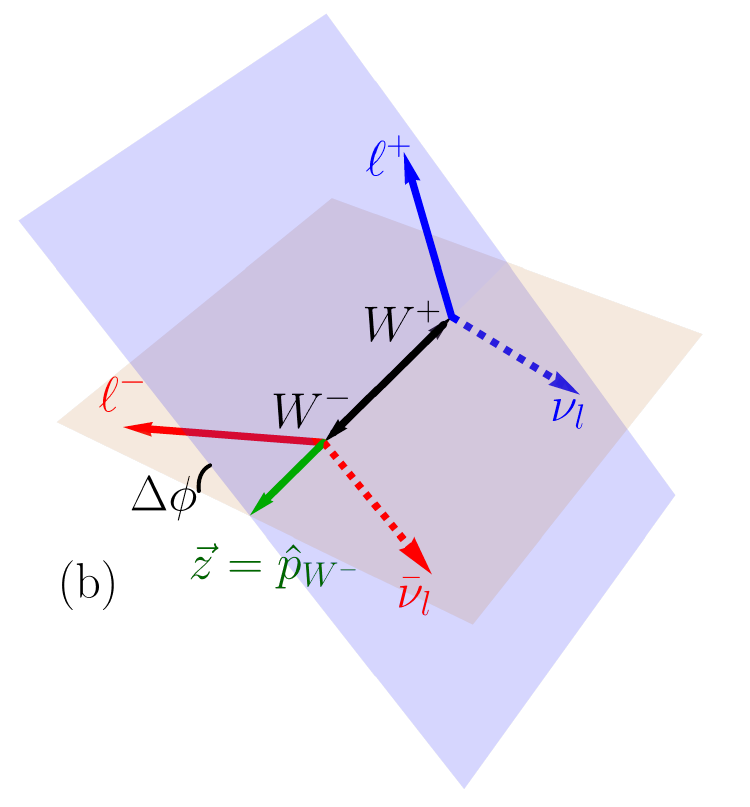}
  \caption{Relative azimuthal angle of $\ell^+$ and $\ell^-$ (a) in the rest frame of $W^+$ and $W^-$, and (b) in the c.m. frame of the $W^\pm$ pair, with the rotational axis $\vec z$ as the moving direction of $W^-$ in the c.m.~frame of $W^\pm$ pair.} 
  \label{fig:Frame_dihedral_angle}
\end{figure}

Despite the challenges in reconstructing the complete spin state of a $W^\pm$ pair, it is useful to start with certain entries in the spin correlation matrix using single observables. The azimuthal-angle correlation is a common choice~\cite{Hagiwara:1986vm,Nelson:1986ki,Chang:1993jy,L3:2003tnr}. Taking the leptonic decay mode of a $W^\pm$ pair as an example
\begin{equation} \label{eq:ww}
W^+ W^- \to \ell^+ \nu_\ell\ \ell^- \bar{\nu}_\ell , 
\end{equation} 
the decay angle of the $\ell^+$/$\ell^-$  ($\ell=e,\mu$) in the rest frame of the $W^+$/$W^-$ carries the polarization information of each mother particle. Going beyond the simple opening angle between the two leptons, the relative azimuthal angle encodes additional information in the transverse dimensions.  Defined with respect to a rotational axis, the azimuthal angle between the two planes formed by the two momenta of $\ell^-$ and $\ell^+$ carries the spin information of the decaying parents and thus their interactions. The spin correlation of the $W^\pm$ pair leads to a nontrivial distribution of the relative azimuthal angle $\Delta\phi$ between $\ell^+$ and $\ell^-$~\cite{Duncan:1985vj,Hagiwara:1986vm} 
\begin{equation}
   \frac{1}{\sigma} \frac{d \sigma}{d\Delta\phi} = \frac{1}{2\pi} + A \cos(\Delta\phi) + \tilde{A} \sin(\Delta\phi)+ (\text{higher partial wave terms}) .  
\end{equation}
The terms from higher order partial waves such as $\cos(2\Delta\phi)$ are less relevant as will be shown later.  Moreover, the observable $\Delta\phi$ is also sensitive to CP properties of the interactions of the $W^\pm$ boson.  The sign of $\Delta\phi$ flips under CP transformations and CP-violating effects in the spin correlation of $W^\pm$ leads to non-zero values of $\tilde{A}$, which is zero according to the Standard Model (SM) predictions at leading order.  Therefore, the relative decay azimuthal angle $\Delta\phi$ captures several important spin correlation effects and has been studied in a number of different processes~\cite{L3:2003tnr,DELPHI:2009wdg,ATLAS:2022oge,Rahaman:2019mnz,Rahaman:2021fcz}.  As a spin correlation observable, the distribution of $\Delta\phi$ is basis dependent because the azimuthal angle needs to be defined with respect to a rotational axis $\vec{z}$ as well as the momentum directions of $\ell^\pm$. A common choice of the rotational axis is the momentum direction of the $W^-$ in the center-of-mass (c.m.) frame of the $W^\pm$ pair (or the lab frame at a lepton collider).  With this basis choice, $\Delta\phi$ is a geometric observable, namely the dihedral angle between the two decay planes, defined either in the individual rest frames of $W^+$ and $W^-$ as shown in Fig.~\ref{fig:Frame_dihedral_angle}(a), or equivalently in the c.m.~frame of the $W^\pm$ pair, as in Fig.~\ref{fig:Frame_dihedral_angle}(b). For a general basis choice, $\Delta\phi$ should be defined in the individual rest frames of $W^+$ and $W^-$.

In this work, we provide a quantitative description of the basis dependence of the azimuthal-angle correlation in the decay of a $W^\pm$ pair.  We present the optimization procedure to achieve the maximal azimuthal-angle correlation analytically. We show that the $W^\pm$ boson moving direction is usually far from the optimal choice. For instance, for $W^\pm$ pair production at a 240~GeV $e^+e^-$ collider, the azimuthal-angle correlation defined with respect to the $W^\pm$ momentum direction can be enhanced about 4 times with an optimal basis choice.  With the rotational axis optimized for the SM CP-even distribution, one can measure the azimuthal-angle correlation to a much higher precision.  With the reference axis optimized for the CP-odd observables, one can achieve the optimal sensitivity of constraining the CP-violating spin correlation effects.

\section{Angular Correlations in Decays and its Basis Dependence}

We first review the azimuthal-angle correlation in the decay of a $W^\pm$ pair and show its basis dependence.  Without loss of generality, we start by choosing $\vec{z}=(0,0,1)$ as the reference direction of the rotational axis to define the azimuthal angle.\footnote{In addition to the rotation axis $\vec z$, the absolute value of azimuthal angle also depends on the choice of $\vec x$ direction where $\phi=0$ is defined, but the relative azimuthal angle $\Delta\phi=\phi-\bar\phi$ only depends on the $\vec z$ direction.} Parametrized by an azimuthal angle $\phi$ and a polar angle $\theta$, the momentum of $\ell^-$ in the rest frame of $W^-$ boson is written as
\begin{equation}
    p_{\ell^-}^\mu=\frac{m_{W}}{2}(1,\,\cos\phi \sin\theta,\, \sin\phi \sin\theta,\, \cos\theta).
\end{equation}
Similarly, the momentum of $\ell^+$ in the rest frame of $W^+$ is parametrized as
\begin{equation}
    p_{\ell^+}^\mu=\frac{m_{W}}{2}(1,\,\cos\bar\phi \sin\bar\theta,\, \sin\bar\phi \sin\bar\theta,\, \cos\bar\theta).
\end{equation}
For a $W^\pm$ pair with their spins correlated, the most general decay angle distribution of $\ell^\pm$ is~\cite{Hagiwara:1986vm} (see also Ref.~\cite{Gounaris:1992kp,Mirkes:1994eb,Groote:2012xr,Stirling:2012zt}).
\begin{equation}\label{eq:fullDistribution}
\begin{aligned}
\frac{d\sigma}{d\Omega^+ d\Omega^-}\propto  \mathcal{A}+\{2 \operatorname{Re}[ & \mathcal{B} \cos \phi+\bar{\mathcal{B}} \cos \bar{\phi}+\mathcal{C} \cos 2 \phi+\bar{\mathcal{C}} \cos 2 \bar{\phi} \\
& +\mathcal{D}_{ \pm} \cos (\phi \pm \bar{\phi})+\mathcal{E}_{ \pm} \cos (2 \phi \pm \bar{\phi})+\bar{\mathcal{E}}_{ \pm} \cos (\phi \pm 2 \bar{\phi}) \\
& \left.\left.+\mathcal{G}_{ \pm} \cos (2 \phi \pm 2 \bar{\phi})\right]+(\operatorname{Re} \rightarrow \mathrm{Im}) + (\cos \rightarrow \sin )\right\},
\end{aligned}
\end{equation}
where the coefficients $(\mathcal{A},\mathcal{B},\cdots)$ are calculated from the helicity amplitudes of $W^\pm$ pair production and $W^\pm$ decays, which also depend on the polar angles $\theta$ and $\bar \theta$; see Appendix~\ref{app:distribution} for details.

In this work, we focus on the correlation between the azimuthal angles of $\ell^+$ and $\ell^-$.  After integrating Eq.~\eqref{eq:fullDistribution} over $\theta$, $\bar\theta$, and $\phi+\bar\phi$, we obtain the distribution with respect to the relative azimuthal angle $\Delta\phi= \phi-\bar\phi$,
\begin{align}
    \frac{1}{\sigma}\frac{d\sigma}{d\Delta\phi} = \frac{1}{2\pi} &+  
    A \cos(\Delta\phi) +
    \tilde{A} \sin(\Delta\phi) +B \cos(2\Delta\phi) +
    \tilde{B} \sin(2\Delta\phi),
    \label{eq:dihedral_distribution}
\end{align}
with the coefficients $A$, $\tilde{A}$, $B$, and $\tilde{B}$ calculated as
\begin{align}
    \label{eq:AfromDensityMatrix}
    A&=\frac{9\pi}{128}\operatorname{Re}\left[\left(\mathscr{P}_{\,0\,0}^{- -} +\mathscr{P}_{0 +}^{- 0} \right)  +\left(\mathscr{P}_{+0}^{0 -} +\mathscr{P}_{++}^{0\,0} \right) \right] , \\
    \label{eq:AtildefromDensityMatrix}
    \tilde{A}&=\frac{9\pi}{128}\operatorname{Im}\left[\left(\mathscr{P}_{\,0\,0}^{- -} +\mathscr{P}_{0 +}^{- 0} \right)  +\left(\mathscr{P}_{+0}^{0 -} +\mathscr{P}_{++}^{0\,0} \right) \right] , \\
    B&= \frac{1}{4\pi} \mathrm{Re}\mathscr{P}^{--}_{++}, \\
    \tilde{B}&= \frac{1}{4\pi} \mathrm{Im}\mathscr{P}^{--}_{++}, 
  \label{eq:densitymatrixDef}
\end{align}
where the spin density matrix and the helicity amplitudes, respectively, are
\begin{equation}
\mathscr{P}^{\lambda\bar\lambda}_{\lambda'\bar\lambda'} = \frac{\overline{\sum}_{\sigma\bar\sigma}\mathcal{M}_{\sigma\bar\sigma,\lambda\bar\lambda} \left(\mathcal{M}_{\sigma\bar\sigma,\lambda'\bar\lambda'}\right)^*}{\overline{\sum}_{\sigma\bar\sigma,\lambda\bar\lambda} |\mathcal{M}_{\sigma\bar\sigma,\lambda\bar\lambda}|^2} ,
\quad\quad
\mathcal{M}_{\sigma\bar\sigma,\lambda\bar\lambda}\equiv\mathcal{M}(e^-_\sigma e^+_{\bar \sigma} \to W^-_{\lambda} W^+_{\bar \lambda}).
\end{equation}
Here, $\sigma = \pm 1$ and $\bar\sigma=\pm 1$ denote the polarizations of the initial $e^+ e^-$ state, and $\lambda = 0,\pm 1$ and $\bar\lambda=0,\pm 1$ denote the helicity of $W^\pm$.

\begin{figure}
	\centering
    \includegraphics[width=0.35\linewidth]{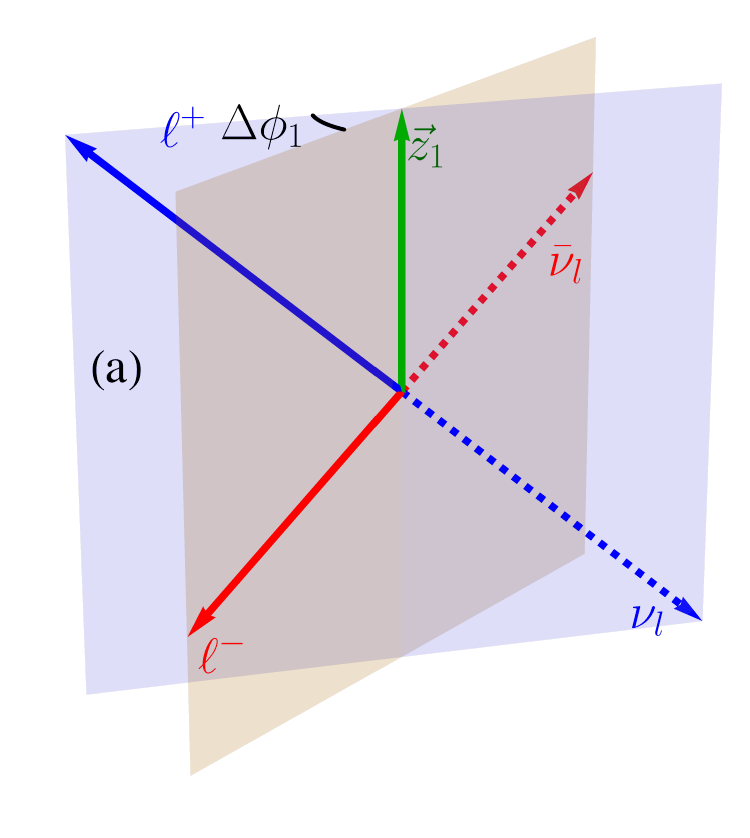}
    \includegraphics[width=0.35\linewidth]{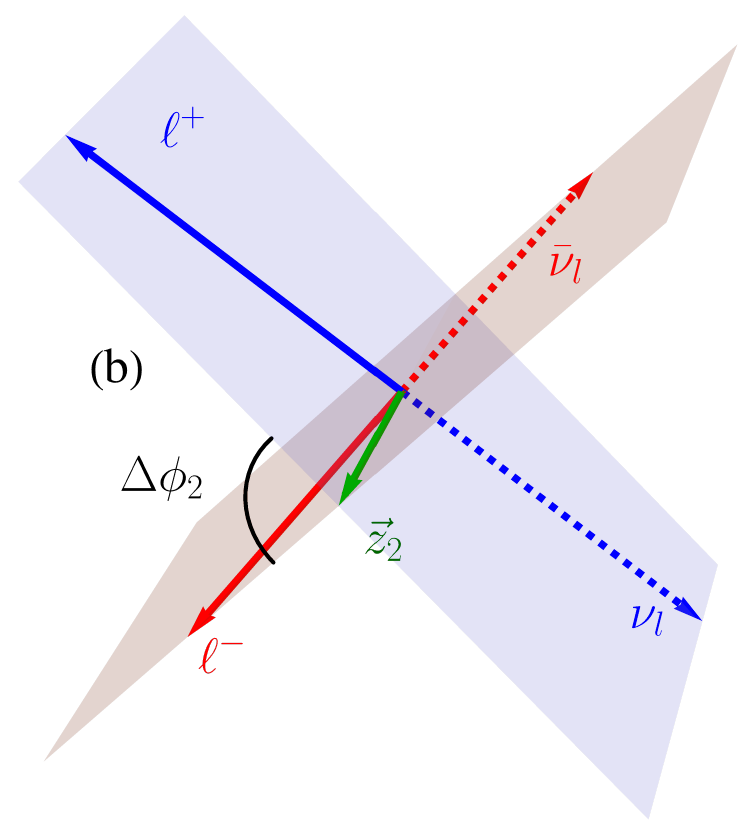}
	\caption{Schematic diagram illustrating the relative azimuthal angle of $\ell^+$ and $\ell^-$ with respect to different rotational $\vec{z}$ direction.  
    The moving directions of $\ell^+$ and $\nu_\ell$ ($\ell^-$ and $\bar \nu_{\ell}$) are defined in the rest frame of $W^+$ ($W^-$). }
	\label{fig:dihedral_angle}
\end{figure}

The distribution of Eq.~\eqref{eq:dihedral_distribution} has been measured at both lepton and hadron colliders~\cite{L3:2003tnr,DELPHI:2009wdg,ATLAS:2022oge} to study the spin correlations of vector boson pairs, in the common helicity basis.  We emphasize that choosing a different rotational axis $\vec{z}$ would lead to a different observable $\Delta\phi$, and thus a different distribution of the angular correlation.  This is illustrated in Fig.~\ref{fig:dihedral_angle}, where the relative azimuthal angle $\Delta\phi$ is defined with respect to different rotational axis $\vec z$. 

\begin{figure}
  \centering
  \includegraphics[scale=0.75]{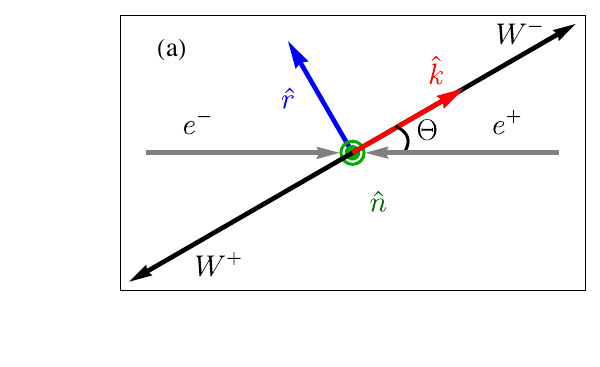}
  \qquad
  \includegraphics[scale=0.75]{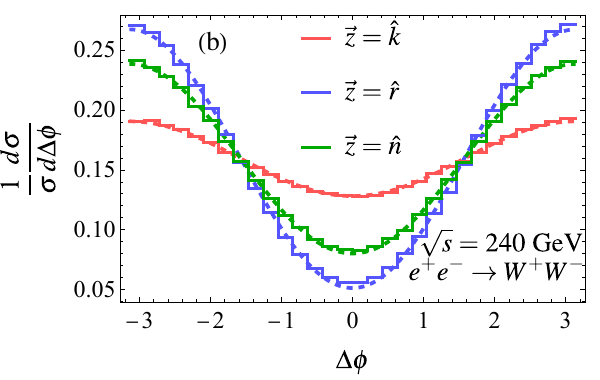}
  \caption{
    (a) The definition  of the helicity basis $(\hat{r},\hat{n},\hat{k})$, 
    (b) $\Delta\phi$ distribution in the decay of $W^+W^-$ produced from the SM process $e^+e^-\to W^+W^-$, where the dotted curves represent theoretical calculations of the leading term in the $\Delta\phi$ distribution and the solid histogram lines represents the $\Delta\phi$ distribution reconstructed from a parton-level \textsc{Madgraph}~\cite{madgraph} simulation.}
    \label{fig:illustration}
\end{figure}

To give a first impression of the basis dependence of Eq.~\eqref{eq:dihedral_distribution}, we calculate the production and decay of a $W^\pm$ pair at a 240 GeV $e^+e^-$ collider, and construct the corresponding $\Delta\phi$ distributions of the decay products of $W^\pm$ with respect to three different directions  $(\hat{r},\hat{n},\hat{k})$ from the commonly used coordinate system known as the helicity basis.  In this basis,
\begin{equation}
\hat{k}=\hat{p}_{W^-},
\qquad\qquad
\hat{r}=\frac{\hat{k} \cos\Theta - \hat{p}_{e^-}}{\sin\Theta},
\qquad\qquad
\hat{n}=\hat{k}\times\hat{r},
\end{equation}
where $\hat p_{e^-} / \hat p_{W^-}$ is the normalized three-momentum direction of $e^-/W^-$, and $\Theta$ is the scattering angle between $e^-$ and $W^-$, as shown in Fig.~\ref{fig:illustration}(a).  The $\Delta\phi$ distributions with respect to these three rotational axes are shown in Fig.~\ref{fig:illustration}(b), and we see that the distribution exhibits a strong dependence on the choice of the rotational axis.  This offers an opportunity to identify a reference rotational axis to optimize the azimuthal-angle correlation. 
The dashed curves in Fig.~\ref{fig:illustration}(b) only include the leading term in the $\Delta\phi$ distribution.  We find that the higher-order terms, $\cos(2\Delta\phi)$ and $\sin(2\Delta\phi)$, are negligible and we only focus on the leading terms hereafter. 

It is interesting that the common choice $\vec z=\hat{k}$ leads to the weakest azimuthal-angle correlation.  This can be qualitatively understood as follows. With the left-handed SM coupling, the outgoing charged leptons are preferably moving parallel (or antiparallel) to the parent momentum $\hat k$, but in opposite directions. This configuration leads to a maximum separation $\Delta\phi \approx \pm\pi$ if azimuthal angles are defined with respect to the perpendicular axis $\hat r$ (or $\hat n$).  As shown in Fig.~\ref{fig:illustration}(b) and to be discussed further, it appears that the axis $\hat r$ in the scattering plane is nearly optimal for the azimuthal-angle correlation.

The axis dependence of the azimuthal-angle correlation shown above can be quantitatively described by formulating the coefficients $A$ and $\tilde{A}$ in Eq.~\eqref{eq:dihedral_distribution} as a function of the rotational axis $\vec z$.  For this, we first consider a Clebsch–Gordan decomposition of the production spin density matrix,
\begin{equation}
\label{eq:Cijdef}
    C_{ij} = \frac{1}{4} \sum_{\lambda,\lambda',\bar\lambda,\bar\lambda'}  \mathscr{P}^{\lambda\bar\lambda}_{\lambda'\bar\lambda'} S^i_{\lambda\lambda'} \bar S^j_{\bar\lambda\bar\lambda'} = \frac{\left\langle  S_i \otimes \bar S_j \right \rangle}{4},
\end{equation}
where $S_i$ and $\bar S_j$ are the angular momentum operators of $W^-$ and $W^+$, respectively; see Appendix~\ref{app:densitymatrix} for details.  Then $C_{ij} \sim \langle S_i\otimes \bar S_j \rangle$ is the spin correlation matrix of the $W^\pm$ pair,\footnote{The complete spin correlation matrix of $W^\pm$ pair is a $8\times 8$ matrix $\braket{\lambda^a\otimes \lambda^b}$ with $\lambda^a$ the Gell-Mann matrices; see Ref.~\cite{Bertlmann:2008jgt} for general discussion and, \textit{e.g.}, Ref.~\cite{Barr:2021zcp} for applications to a vector boson pair. The spin correlation matrix $C_{ij}$ we defined here is a $3\times 3$ submatrix of it; see Refs.~\cite{Rahaman:2021fcz,Bi:2023uop} for its relation to the complete $8\times 8$ matrix.} which transforms as $C \to R^T CR $ under the basis transformation $\vec z\to R \vec z$.  Combining Eq.~\eqref{eq:Cijdef} with Eqs.~\eqref{eq:AfromDensityMatrix} and \eqref{eq:AtildefromDensityMatrix}, the coefficients $A$ and $\tilde{A}$ can be expressed in terms of the spin correlation matrix as
\begin{equation}
    A=-\frac{9\pi}{64}\left(C_{ii}-C_{33}\right),
    \qquad\qquad 
    \tilde{A}=-\frac{9\pi}{64}\varepsilon_{ij3}C_{ij},
\end{equation}
where the $3^{\rm rd}$ direction is $\vec{z}$,  $\varepsilon_{ijk}$ is the antisymmetric tensor, and repeated indices are summed over.  With the basis transformation rule of the correlation matrix under $\vec z\to R\vec z$,  the dependence of $A$ and $\tilde{A}$ on the axis $\vec z$ can be expressed in an explicitly SO(3) invariant way 
\begin{equation}
    A(\vec{\vspace{1pt}z}\hspace{1pt} )=-\frac{9\pi}{64}\left(C_{ii}-C_{ij}z_iz_j\right),
    \qquad \qquad 
    \tilde{A}(\vec{\vspace{1pt}z}\hspace{1pt})=-\frac{9\pi}{64}\varepsilon_{ijk}C_{ij}z_{k}.
\label{eq:AandAtilde}
\end{equation}
This leads to an intuitive connection between the spin correlation matrix and the decay azimuthal correlation, \textit{e.g.}, the strength of the $\cos(\Delta\phi)$ modulation along $\vec{z}$ direction is directly related to the components of the $W^\pm$ spin correlation matrix along the $\vec{z}$ direction $\braket{S_z\otimes \bar S_z}$.

\section{Optimizing the Azimuthal-angle Correlations in the decay of $W^\pm$ pair}
\label{sec:optimization}

The leading-order differential distribution has two oscillatory terms: a $\cos(\Delta\phi)$ term with a coefficient $A$ and a $\sin(\Delta\phi)$ term with a coefficient $\tilde{A}$. Each of these terms has a distinct basis dependence on the axis $\vec{z}$, as seen in Eq.~\eqref{eq:AandAtilde}, meaning that one choice of $\vec{z}$ will maximize $A$ and a separate choice of $\vec{z}$ will maximize $\tilde{A}$.  Depending on the physics we are interested in, we can choose which coefficient we optimize.  In particular, the $\cos(\Delta\phi)$ and $\sin(\Delta\phi)$ modulations are, respectively, related to CP-even and CP-odd interactions.  Under a CP transformation, the angles transform as
\begin{equation}
  \begin{aligned}   
  \left(\theta,\phi,\bar{\theta},\bar{\phi}\right)&\xrightarrow{\rm{CP}}\left(\pi-\bar{\theta},\pi+\bar{\phi},\pi-\theta,\pi+\phi\right),\\
    \Delta\phi &\xrightarrow{\rm{CP}} -\Delta\phi,
  \end{aligned}
\end{equation}
therefore, the $\sin(\Delta\phi)$ term flips its sign while the $\cos(\Delta\phi)$ term is invariant.  Thus, the coefficient $A$ parametrizes CP-even interactions while $\tilde{A}$ parametrizes CP-odd interactions.
The spin correlation matrix defined in Eq.~\eqref{eq:Cijdef} transforms as $C_{ij}\to C_{ji}$ under a CP transformation.
From Eq.~\eqref{eq:AandAtilde}, we see that $A$ only receives contributions from the symmetric part of $C_{ij}$ which is CP-even, while $\tilde{A}$ only receives contributions from the antisymmetric part of $C_{ij}$ which is CP-odd.

In the SM at leading order, $W^+ W^-$ production is CP-invariant, therefore to best characterize the SM, the coefficient $A$ should be optimized.  Meanwhile, CP-violating new physics effects on the spin correlations of $W^\pm$ are manifested in the coefficient $\tilde{A}$.  In Sec.~\ref{sec:optA} we show the optimization of $A$ and in Sec.~\ref{sec:optAtilde} we show the optimization of $\tilde{A}$.

\subsection{Optimizing the Azimuthal-Angle Correlation: CP-Even Interactions}
\label{sec:optA}

\begin{figure}
	\centering	
    \includegraphics[width=0.45\linewidth]{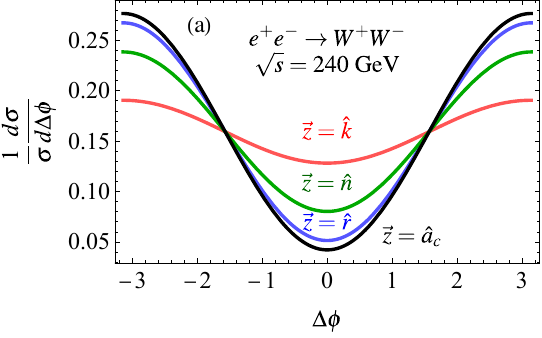}
    \qquad
     \includegraphics[width=0.45\linewidth]{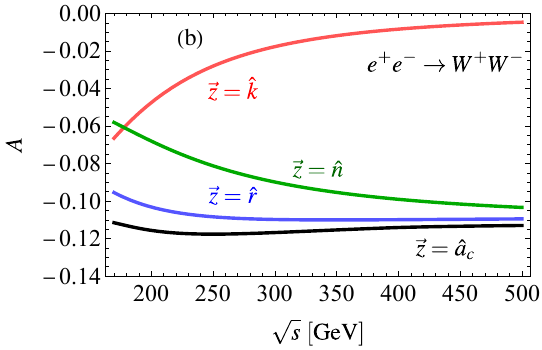}
	\caption{The SM prediction of (a) the distribution of relative azimuthal angle of $\ell^+$ and $\ell^-$ in $W^\pm$ pair production with respect to $\hat r,\ \hat n$, $\hat k$, and the optimal direction $\hat a_c$, and (b) the value of $A$ with respect to different rotational axes as a function of collision energy.}
	\label{fig:distribution}
\end{figure}

We first focus on the leading-order SM prediction of the azimuthal-angle correlation, with the magnitude of the $\cos(\Delta\phi)$ modulation given by $|A|$ in Eq.~\eqref{eq:AandAtilde}.  The basis dependence of $|A| \propto \tr(C)-\vec z\cdot C \cdot \vec z $ is determined by $\vec z \cdot C \cdot \vec z$, \textit{i.e.}, the diagonal entry of $C_{ij}$ along the $\vec{z}$ direction.  As the diagonal entry of a symmetric matrix is bounded by its largest and smallest eigenvalues, the maximum of $|A|$ is achieved when $\vec z$ is the eigenvector, denoted as $\hat a_c$,\footnote{The subscript $c$ refers to the cosine term. The full analytical expression of $\hat a_c$ is lengthy and not illuminating. We choose only present the numerical results in the following.} of $C_{ij}$ with the smallest eigenvalue.\footnote{In the SM, $\tr(C)-\vec{z}\cdot C \cdot \vec{z}$ is always positive so that its maximum value is achieved when $\vec{z}\cdot C \cdot \vec{z}$  is minimal.}  Therefore, to investigate the $\cos(\Delta\phi)$ distribution predicted by the SM, it is optimal to choose $\hat{a}_c$ as the rotational axis to define the azimuthal angle.

We calculate the spin correlation matrix $C_{ij}$ defined in Eq.~\eqref{eq:Cijdef} from the helicity amplitude of $e^+e^- \to W^+ W^-$ in the SM and we obtain the eigenvector $\hat a_c$ of $C_{ij}$ with smallest eigenvalue.  Then we define the azimuthal angle $\phi$ and $\bar \phi$ of the decay products of $W^-$ and $W^+$ with respect to $\hat a_c$ and show the $\Delta\phi$ distribution in Fig.~\ref{fig:distribution}(a).  Compared to the commonly used $\hat{k}$ direction to define the azimuthal angle, the optimal axis yields a much larger correlation, with about a 4 times larger coefficient $A$, as shown in Fig.~\ref{fig:distribution}(a).  Consequently, a significantly smaller relative uncertainty of the azimuthal-angle correlation is obtained with the optimal axis choice. 
In practice, as shown in Fig.~\ref{fig:distribution}(a), the azimuthal-angle correlation defined with respect to the $\hat{r}$ direction is close to its maximal value.  For practical use, the direction $\hat r$ is convenient and approximately optimal when used to measure the azimuthal-angle correlation.

Figure~\ref{fig:distribution}(b) shows the calculated coefficient $A$ for the axis choice $\hat{k}$, $\hat{n}$, $\hat{r}$, and the optimal axis $\hat{a}_c$ as a function of c.m. energy.  The azimuthal-angle correlation with respect to the $\hat{k}$ direction decreases with energy.  
This is because the $W^\pm$ production becomes more collinear with the initial beam at higher energies, therefore the system exhibits an approximate azimuthal rotation symmetry with respect to the moving direction $\hat{k}$ of $W^-$, and leads to a flat distribution of $\Delta\phi$ and little correlation.

The above optimal basis is obtained at leading order.  A complete higher-order treatment is beyond the scope of this work.
However, since the NLO electroweak correction to the spin observables of $e^+e^-\to W^+W^-$ is within 10\%~\cite{Denner:2005fg}, we estimate that the optimal basis at tree level still serves as the approximately optimal basis at higher orders, especially if any additional high $p_T$ objects in the events are vetoed to explicitly maintain the leading order kinematics.

\subsection{Optimizing the Azimuthal-Angle Correlation: CP-Odd Interactions}
\label{sec:optAtilde}

The above optimization procedure is equally applicable to the $\sin(\Delta\phi)$ modulation in the distribution.   From Eq.~\eqref{eq:AandAtilde}, the magnitude of the $\sin(\Delta\phi)$ modulation is maximal when we choose the reference axis $z_k \propto \varepsilon_{ijk} C_{ij}$.  As the $W^\pm$ pair production in the SM is CP-conserving, this optimal basis choice for the $\sin(\Delta\phi)$ modulation serves to maximize the sensitivity of probing CP-violating spin correlations introduced by new physics effects.

As an illustration, we work in the SMEFT framework where there are only two CP-violating dimension-six operators that modify the vector boson self-interactions~\cite{Degrande:2012wf}
\begin{equation}  
\label{operator:CPVc3}
\mathcal{O}_{\widetilde{W}WW} =\tr\left[\widetilde{W}_{\mu\nu}W^{\nu\rho}W_{\rho}^{\mu}\right],
\qquad \qquad 
\mathcal{O}_{\widetilde{W}} =\left(D_{\mu}\Phi\right)^{\dagger}\widetilde{W}^{\mu\nu}\left(D_{\nu}\Phi\right),
\end{equation}
where $\Phi$ is the Higgs doublet field and
\begin{align}
D_\mu & =\partial_\mu+\frac{i}{2} g \tau^I W_\mu^I+\frac{i}{2} g^{\prime} B_\mu, \\
W_{\mu \nu} & =\frac{i}{2} g \tau^I\left(\partial_\mu W_\nu^I-\partial_\nu W_\mu^I+g \epsilon_{I J K} W_\mu^J W_\nu^K\right).
\end{align}
The effective Lagrangian is written as
\begin{equation}
    \mathcal{L}_{\rm SMEFT}=\mathcal{L}_{\rm SM} +  \cwww \frac{\mathcal{O}_{\widetilde{W}WW}}{\Lambda^2} + \cw \frac{\mathcal{O}_{\widetilde{W}}}{\Lambda^2},
\end{equation}
where $\Lambda$ is the new physics scale and $\cw$ and $\cwww$ are dimensionless Wilson coefficients.

The leading-order contributions of $\cw$ and $\cwww$ in $W^\pm$ pair production are the interferences between the new physics and the SM amplitude, which does not arise in the total cross section but in the spin correlation between the $W^\pm$ pair.   The spin correlation matrix of the $W^\pm$ pair, including the new physics contribution, is
\begin{equation}
    C_{ij} = C_{ij}^{\rm SM} + C_{ij}^{\rm int} + \mathcal{O}(c_{\widetilde W}^2, c_{\widetilde WWW}^2),
\end{equation}
where $C_{ij}^{\rm SM}$ is the symmetric part of the spin correlation matrix given by the SM prediction, and $C_{ij}^{\rm int}$ is the antisymmetric part of the spin correlation matrix, which is given by the interference effect and depends linearly on $c_{\widetilde{W}}$ and $c_{\widetilde WWW}$.  From Eqs.~\eqref{eq:densitymatrixDef} and~\eqref{eq:Cijdef}, $C_{ij}^{\rm int}$ is calculated as
\begin{equation}
    C_{ij}^{\rm int} = \frac{1}{4} \sum_{\lambda,\lambda',\bar\lambda,\bar\lambda'}   S^i_{\lambda\lambda'} \bar S^j_{\bar\lambda\bar\lambda'} \frac{\sum_{\sigma\bar\sigma}\mathcal{M}_{\sigma\bar\sigma,\lambda\bar\lambda}^{\rm SM} \left(\mathcal{M}_{\sigma\bar\sigma,\lambda'\bar\lambda'}^{\rm EFT}\right)^*}{\sum_{\sigma\bar\sigma,\lambda\bar\lambda} |\mathcal{M}_{\sigma\bar\sigma,\lambda\bar\lambda}^{\rm SM}|^2} + {\rm h.c.},
\end{equation}
and therefore linearly related to the amplitudes of the new physics contribution and the corresponding Wilson coefficients.

\begin{figure}
  \centering
  \includegraphics[width=0.45\linewidth]{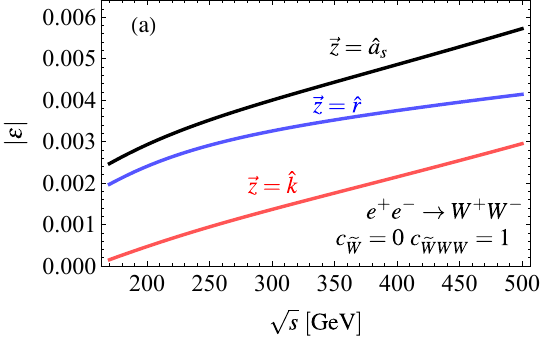}
  \qquad
  \includegraphics[width=0.45\linewidth]{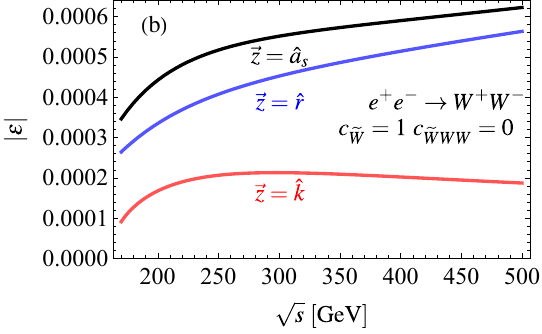}
  \caption{The value of $|\epsilon|$ with respect to different reference axes as a function of collision energy. (a)  $c_{\widetilde{W}}=0$ and $c_{\widetilde{W}WW}=1$. (b) $c_{\widetilde{W}}=1$ and $c_{\widetilde{W}WW}=0$.
    The new physics scale for operators is $\Lambda=1\tev$.}
  \label{fig:AtidleWithEnergy}
\end{figure}

As shown in Eq.~\eqref{eq:AandAtilde}, the antisymmetric part of the spin correlation matrix, namely $C_{ij}^{\rm int}$, leads to the CP-odd $\sin(\Delta\phi)$ term in the azimuthal-angle distribution of $W^\pm$ pair decay, with the magnitude $\tilde{A}\propto \varepsilon_{ijk} C^{\rm int}_{ij} z_k$.  To extract this CP-odd distribution, a frequently-employed CP observable is the azimuthal asymmetry~\cite{Chang:1993jy,Han:2009ra,Rahaman:2019mnz}
\begin{equation}
    \epsilon \equiv \frac{N(\Delta\phi>0)-N(\Delta\phi<0)}{N(\Delta\phi>0)+N(\Delta\phi<0)} 
    =\frac{1}{\sigma}\left(\int_0^\pi \frac{d\sigma}{d\Delta\phi} d\Delta\phi-\int_{-\pi}^0 \frac{d\sigma}{d\Delta\phi} d\Delta\phi\right) = 4\tilde{A},
\end{equation} 
Therefore, after introducing the CP-odd operators, the asymmetry $\epsilon $ is induced by the antisymmetric part of spin correlation matrix $C_{ij}^{\rm int}$
\begin{equation}
    \epsilon =- \frac{9\pi}{16} \varepsilon_{ijk} C_{ij} z_{k} = -\frac{9\pi}{16} \varepsilon_{ijk} C_{ij}^{\rm int} z_{k}.
\end{equation}
We calculate the asymmetry $\epsilon$ again  with respect to three different axis: $\hat k$, $\hat r$, as well as the optimal axis $\hat a_s$ defined as\footnote{The subscript $s$ refers to the sine term.}
\begin{equation}
(\hat{a}_s)_k \propto \epsilon_{ijk}C_{ij}.
\end{equation}
We show the CP asymmetry $|\epsilon|$ versus the $e^+e^-$ c.m.~energy with respect to different axes in Fig.~\ref{fig:AtidleWithEnergy}(a) for $c_{\widetilde{W}}=0$ and $c_{\widetilde WWW}=1$, and Fig.~\ref{fig:AtidleWithEnergy}(b) for  $c_{\widetilde{W}}=1$ and $c_{\widetilde WWW}=0$.  As expected, the operator $\mathcal{O}_{\widetilde{W}WW}$ posseses a stronger energy-dependence than $\mathcal{O}_{\widetilde{W}}$. When the azimuthal angle is defined with respect to the optimal axis $\hat a_s$, we maximize the $\sin(\Delta\phi)$ distribution and thus obtain the maximal azimuthal asymmetry $\epsilon$. We therefore achieve the strongest sensitivity. 

\begin{figure}
	\centering
    \includegraphics[scale=0.7]{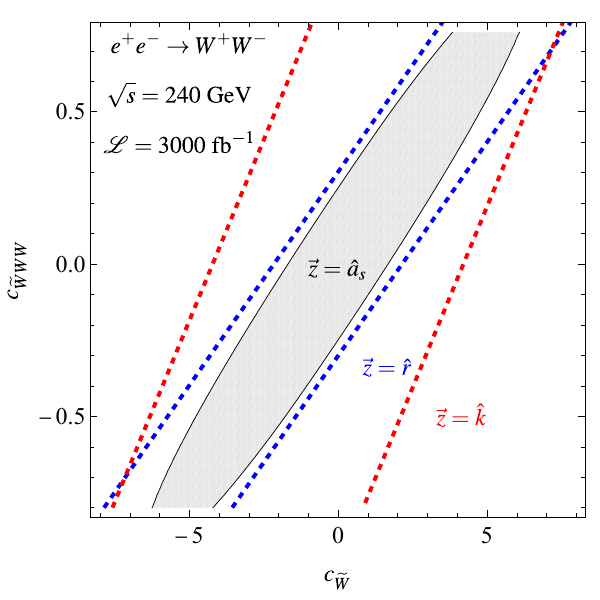}
	\caption{The allowed parameter region at $2 \sigma$ confidence level from the CP asymmetry observable $\epsilon$ reconstructed with respect to different axes, in the dilepton decay channel of $W^\pm$ pair production at a $240$ GeV electron-positron collider.
    The new physics scale for operators is $\Lambda=1\tev$.
    }
	\label{fig:2sigma}
\end{figure}

With a sufficient number of events $N_{\rm tot}$, the statistical uncertainty of $\epsilon$ can be estimated by $\delta\epsilon = 1/\sqrt{N_{\rm tot}}$ and the constraints on CP-odd operators are calculated accordingly. 
In Fig.~\ref{fig:2sigma}, we show the $2\sigma$ exclusion limit of $c_{\widetilde{W}}$ and $c_{\widetilde{W}WW}$ from the decay azimuthal asymmetry $\epsilon$ of a $W^\pm$ pair produced at a $240$ GeV $e^+e^-$ collider.  The inner shaded region by the solid lines presents the most sensitive results with the $2\sigma$ bound using the optimal rotational axis $\vec{z}=\hat{a}_s$, which is about three times smaller than the constraint given by using $\vec{z}=\hat{k}$.  With the single observable $\epsilon$ in the optimal basis, we find that the constraint on $c_{\widetilde{W}}$ and $c_{\widetilde{W}WW}$ is the same order of magnitude as the global fit results with machine learning techniques shown in Ref.~\cite{Subba:2022czw}. We conclude that the most relevant CP-violating effects can be captured by the $\sin(\Delta\phi)$ distribution in the optimal basis.

\section{Conclusions}

Spin correlation observables are generally coordinate basis dependent for particles produced at colliders.  In this work, by performing a Clebsch–Gordan decomposition of the spin density matrix, we quantitatively described the basis dependence of the spin correlation between $W^\pm$ pairs focusing on the azimuthal-angle correlation in their decays.  Utilizing this basis dependence, we further provided the optimal basis to define the relative azimuthal angle $\Delta\phi$ to achieve the maximal spin correlation. 

With the optimization procedure we developed, one can choose to maximize different modulations in the distribution of the relative azimuthal angle $\Delta\phi$ for different physics purposes.  For a better measurement of the SM prediction, one can maximize the CP-even $\cos(\Delta\phi)$ modulation to yield the largest possible value of the coefficient $A$.  The optimal basis choice yields a correlation about 4 times larger compared to the conventional choice along the $W^\pm$ moving direction, leading to the optimal sensitivity of measuring the SM prediction of azimuthal-angle correlation.  To study the CP-odd component of $W^\pm$ pair spin correlation from new physics effects with a single CP asymmetry observable $\epsilon$,  one can maximize the $\sin(\Delta\phi)$ modulation and achieve the optimal constraint on the CP-odd operator.  In addition, the optimal direction is close to the direction $\hat r$ perpendicular to the $W^\pm$ momentum in the $e^+e^-\to W^+ W^-$ scattering plane for both CP-even and CP-odd interactions, making our suggestion of using the $\hat r$ direction in lieu of the mathematically optimal one more convenient to apply in realistic studies.

In our presentation, we considered the leptonic decay channels as in Eq.~\eqref{eq:ww} with $\ell=e,\mu$. In practice, there is a two-fold ambiguity when solving the neutrino momentum from the $W$ mass constraint~\cite{Hagiwara:1986vm}. This may lead to some degree of degradation in the sensitivity results, although our optimization procedure is still valid. On the other hand, one can also generalize this analysis to the semi-leptonic channel $W^+W^-\to \ell^\pm\nu\ q \bar q'$.  In this channel the branching fraction is larger and kinematic reconstruction is easier. The drawback is the need for flavor tagging or charge tagging for one of the hadronic jets. 

The quantitative description of the basis dependence and the optimization procedure in this work applies equally to the spin correlation of a $ZZ$ pair and the $ZW^\pm$ final state. Moreover, the Clebsch–Gordan decomposition characterizes how the spin correlation transforms under an SO(3) basis transformation, which provides a general way to quantitatively describe the coordinate basis dependence of the spin correlations in any system.

With the upcoming LHC Run 3 and the HL-LHC runs, as well as the planning of FCC-ee, one would expect to accumulate a large number of events of gauge boson pairs. Our procedure surely will provide a new avenue to effectively study the rich physics associated with gauge boson pairs, and look for subtle CP violation effects from physics beyond the Standard Model.

Finally, we note that there is potential for future work when the $W^+ W^-$ system is treated as a quantum state (see also Ref.~\cite{Grabarczyk:2024wnk}).  While our work optimizes a single entry of the spin correlation matrix, analogous to Refs.~\cite{Parke:1996pr,Mahlon:1997uc,Uwer:2004vp} in the $t\bar{t}$ system, a natural follow-up would be to perform optimization for quantum entanglement or Bell nonlocality, as was done in Refs.~\cite{Cheng:2023qmz,Cheng:2024btk} for the $t\bar{t}$ system.

\begin{acknowledgments}
This work was supported in part by the U.S.~Department of Energy under grant No.~DE-SC0007914 and in part by the Pitt PACC. Yi-Jing Fang is supported in part by the National Science Foundation of China under Grants No.~12235001. ML is also supported by the U.S.~National Science Foundation under grant No.~PHY-2112829. TH would also like to thank the hospitality extended by the Kavli Institute for Theoretical Physics (KITP), which was supported by grant NSF PHY-2309135. 
\end{acknowledgments}

\appendix

\section{The Production Density Matrix and Spin Correlation Matrix of a $W^\pm$ Boson Pair}
\label{app:densitymatrix}

At a given phase space point, the spin density matrix of $W^+ W^-$ produced from $e^+ e^-$ annihilation is
\begin{equation}
    \mathscr{P}^{\lambda\bar\lambda}_{\lambda'\bar\lambda'}= \frac{\sum_{\sigma\bar\sigma}\mathcal{M}_{\lambda\bar\lambda}^{\sigma\bar\sigma} \left(\mathcal{M}_{\lambda'\bar\lambda'}^{\sigma\bar\sigma}\right)^*}{\sum_{\sigma\bar\sigma,\lambda\bar\lambda} |\mathcal{M}_{\lambda\bar\lambda}^{\sigma\bar\sigma}|^2} ,
\end{equation}
where the matrix element is
\begin{equation}
 \mathcal{M}_{\lambda \bar\lambda}^{\sigma\bar\sigma} \equiv \bra{W^-(k_1,\lambda)W^+(k_2,\bar \lambda)}\mathcal{T}\ket{e^-(p_1,\sigma)e^+(p_2,\bar \sigma)},
\end{equation}
and $\mathcal{T}$ is the transition matrix. $p_i/k_i$ is the momentum of the initial/final state; $\sigma = \pm 1$ and $\bar\sigma=\pm 1$ denote the polarizations of the initial $e^+ e^-$ state;  $\lambda = 0,\pm 1$ and $\bar\lambda=0,\pm 1$ denote the helicity of $W^\pm$.

The helicity index $\lambda$ of the $W^\pm$ boson is defined according to the polarization vector.  In Ref.~\cite{Hagiwara:1986vm}, they are chosen as
\begin{equation}
\epsilon_+= \frac{1}{\sqrt{2}}(0,-1,-i,0),
\qquad\qquad
\epsilon_0= (0,0,0,1),
\qquad\qquad
\epsilon_-= \frac{1}{\sqrt{2}}(0,1,-i,0),
\end{equation}
for $W^-$ in its rest frame and
\begin{equation}
\bar\epsilon_+= \frac{1}{\sqrt{2}}(0,1,-i,0),
\qquad\qquad
\bar\epsilon_0= (0,0,0,1),
\qquad\qquad
\bar\epsilon_-= \frac{1}{\sqrt{2}}(0,-1,-i,0),
\end{equation}
for $W^+$ in its rest frame, where we have chosen the $W^-$ moving direction as $\vec z$ direction for simplicity.  Note that the helicity of the $W^+$ is opposite to its spin.

Consider a new direction $\vec z^{\,\prime} = R\cdot \vec z$ as the spin quantization axis, then the spin basis $\epsilon_\lambda$ transform as $\epsilon_{\lambda} \to U_{\lambda\lambda'} \epsilon_{\lambda'}$ and $\bar{\epsilon}_{\lambda} \to \bar{U}_{\lambda\lambda'} \bar{\epsilon}_{\lambda'}$, where $U$ and $\bar{U}$ are 3-dimensional unitary matrices.
The unitarity transformation $U$ of the spin basis can be related to the rotation $R$ with the spin operators. The spin operators of $W^-$ expressed in the basis $(\epsilon_+,\epsilon_0,\epsilon_-)$ are
\begin{equation}
    S_x=\frac{1}{\sqrt{2}}\begin{pmatrix}
        0 & 1 & 0 \\
        1 & 0 & 1 \\
        0 & 1 & 0 \\
    \end{pmatrix},
    \qquad\qquad
    S_y=\frac{1}{\sqrt{2}}\begin{pmatrix}
        0 & -i & 0 \\
        i & 0 & -i \\
        0 & i & 0 \\
    \end{pmatrix},
    \qquad\qquad
    S_z=\begin{pmatrix}
        1 & 0 & 0 \\
        0 & 0 & 0 \\
        0 & 0 & -1 \\
    \end{pmatrix}.
\end{equation}
The helicity direction of $W^+$ is defined according to $-\vec{z}$ direction, opposite to the helicity definition of $W^-$.  
If we still express the spin operator in the order of helicity, $\textit{i.e.}$, $(\bar{\epsilon}_+,\bar{\epsilon}_0,\bar{\epsilon}_-)$, then we need to flip the indices of the spin operator of $W^-$, with $\bar{S}_x=S_x$,  $\bar{S}_y=-S_y$, and $\bar{S}_z=-S_{z}$.
For each representation of the spin operator, under the basis transformation, we have
\begin{equation}
    S_i \to U S_i U^\dagger = R_{ji} S_j,
    \qquad\qquad
    \bar{S}_i \to \bar{U} \bar{S}_i \bar{U}^\dagger = R_{ji} \bar{S}_j.
\end{equation}
Then we see that the spin correlation matrix
\begin{equation}
    C_{ij} = \frac{1}{4}\sum_{\lambda,\lambda',\bar\lambda,\bar\lambda'}  \mathscr{P}^{\lambda\bar\lambda}_{\lambda'\bar\lambda'} S^i_{\lambda\lambda'} \bar S^j_{\bar\lambda\bar\lambda'}
\end{equation}
transforms as an SO(3) tensor.

If higher dimensional operators are introduced, the production density matrix can be written in 3 terms, 
\begin{equation}
   \mathscr{P}^{\lambda\bar\lambda}_{\lambda'\bar\lambda'}= \mathscr{P}^{\lambda\bar\lambda{(0)}}_{\lambda'\bar\lambda'}+ \mathscr{P}^{\lambda\bar\lambda(1)}_{\lambda'\bar\lambda'}+ \mathscr{P}^{\lambda\bar\lambda(2)}_{\lambda'\bar\lambda'},
\end{equation}
where the SM contribution term $\mathscr{P}^{\lambda\bar\lambda{(0)}}_{\lambda'\bar\lambda'}$, the interference term between the SM and the new physics $\mathscr{P}^{\lambda\bar\lambda{(1)}}_{\lambda'\bar\lambda'}$, and the quadratic term of the new physics contribution $\mathscr{P}^{\lambda\bar\lambda{(2)}}_{\lambda'\bar\lambda'}$, respectively, are
\begin{equation}
    \begin{aligned}
    &\mathscr{P}^{\lambda\bar\lambda{(0)}}_{\lambda'\bar\lambda'}=\frac{1}{\mathcal{N}} \sum_{\sigma\bar\sigma}\mathcal{M}^{\rm SM}_{\sigma\bar\sigma,\lambda\bar\lambda}\big(\mathcal{M}^{\rm SM}_{\sigma\bar\sigma,\lambda'\bar\lambda'}\big)^*,\\   
    &\mathscr{P}^{\lambda\bar\lambda{(1)}}_{\lambda'\bar\lambda'}=\frac{1}{\mathcal{N}} \sum_{\sigma\bar\sigma}\left[\mathcal{M}^{\rm SM}_{\sigma\bar\sigma,\lambda\bar\lambda}\big(\mathcal{M}^{\rm EFT}_{\sigma\bar\sigma,\lambda'\bar\lambda'}\big)^* + \mathcal{M}^{\rm EFT}_{\sigma\bar\sigma,\lambda\bar\lambda}\big(\mathcal{M}^{\rm SM}_{\sigma\bar\sigma,\lambda'\bar\lambda'}\big)^*\right],\\   
    &\mathscr{P}^{\lambda\bar\lambda{(2)}}_{\lambda'\bar\lambda'}=\frac{1}{\mathcal{N}} \sum_{\sigma\bar\sigma}\mathcal{M}^{\rm EFT}_{\sigma\bar\sigma,\lambda\bar\lambda}\big(\mathcal{M}^{\rm EFT}_{\sigma\bar\sigma,\lambda'\bar\lambda'}\big)^*,\\   
    \end{aligned}
\end{equation}
and the normalization factor $\mathcal{N}$ is
\begin{equation}
        \mathcal{N}=\sum_{\sigma\bar\sigma,\lambda\bar\lambda} \left| \mathcal{M}^{\rm SM}_{\sigma\bar\sigma,\lambda\bar\lambda} + \mathcal{M}^{\rm EFT}_{\sigma\bar\sigma,\lambda\bar\lambda}\right|^2=\sum_{\sigma\bar\sigma,\lambda\bar\lambda} \left|\mathcal{M}^{\rm SM}_{\sigma\bar\sigma,\lambda\bar\lambda} \right|^2 +\sum_{\sigma\bar\sigma,\lambda\bar\lambda} \left| \mathcal{M}^{\rm EFT}_{\sigma\bar\sigma,\lambda\bar\lambda}\right|^2.
\end{equation}
Note that the interference term does not contribute to the total cross section
\begin{equation}
        \sum_{\sigma\bar\sigma,\lambda\bar\lambda} \Big[\mathcal{M}^{\rm SM}_{\sigma\bar\sigma,\lambda\bar\lambda}\big(\mathcal{M}^{\rm EFT}_{\sigma\bar\sigma,\lambda\bar\lambda}\big)^* + \mathcal{M}^{\rm EFT}_{\sigma\bar\sigma,\lambda\bar\lambda}\big(\mathcal{M}^{\rm SM}_{\sigma\bar\sigma,\lambda\bar\lambda}\big)^*\Big]=0.
\end{equation}

\section{Decay-Angle Correlations for Final State Leptons}
\label{app:distribution}

The angular distribution of the leptons from the $W^\pm$ decays can be expressed as the product of the production density matrix and the decay density matrix of the $W^\pm$ boson as follows: 
\begin{equation}
    \frac{d\sigma}{d\Omega^+ d\Omega^-}=\frac{1}{N}\mathscr{P}^{\lambda\bar\lambda}_{\lambda'\bar\lambda'}D^{\lambda}_{\lambda'}\bar D^{\bar\lambda}_{\bar\lambda'}.
\end{equation}
Here, $N$ is the normalization factor which is $N=9/(64\pi^2)$, $D^{\lambda}_{\lambda'}$ and $\bar D^{\bar \lambda}_{\bar \lambda'}$ are the normalized decay density matrix of $W^-$ and $W^+$, respectively.

To obtain the distribution, we first define the momenta of the leptons by choosing the helicity basis, shown in Fig.~\ref{fig:illustration}(a) in the rest frame of the $W^-$ and $W^+$ bosons. The four-momenta of leptons in terms of polar angle and azimuthal angle can be expressed as
\begin{equation}
    \begin{aligned}
        p_{\ell^-}^\mu&=\frac{m_{W^-}}{2}(1,\,\cos\phi \sin\theta,\, \sin\phi \sin\theta,\, \cos\theta),\\
        p_{\ell^+}^\mu&=\frac{m_{W^+}}{2}(1,\,\cos\bar\phi \sin\bar\theta,\, \sin\bar\phi \sin\bar\theta,\, \cos\bar\theta).
    \end{aligned}
\end{equation}
In the decay of a $W^\pm$ boson, the final states are angular momentum eigenstates in the $W^\pm$ boson rest frame because there are only left-handed interactions.  Therefore, the decay amplitude can be expressed in the form of Wigner functions~\cite{Hagiwara:1986vm},
\begin{equation}
  \mathcal{M}_{W^-_\lambda} \propto l_{\lambda},
  \qquad
  \mathcal{M}_{W^+_{\bar\lambda}}\propto \bar l_{
       \bar\lambda},
\end{equation}
where 
\begin{equation}
  \left(l_+,l_0,l_-\right)=\left(d_-e^{i\phi}, -d_0, d_+e^{-i\phi}\right),
  \qquad 
  \left(\bar l_+,\bar l_0,\bar l_-\right)=\left(\bar d_-e^{-i\bar \phi}, -\bar d_0, \bar d_+e^{i\bar \phi}\right),
\end{equation}
with 
\begin{equation}
    \begin{aligned}
        & d_{\pm}=\frac{1}{\sqrt{2}}(1\pm\cos{(\theta)}), 
        & \qquad d_0=\sin{(\theta)}, \\
        & \bar d_{\pm}=\frac{1}{\sqrt{2}}(1\pm\cos{(\bar \theta)}), 
        & \qquad \bar d_0=\sin{(\bar \theta)}.\\
    \end{aligned}
\end{equation}
Therefore, the normalized density matrix of $W^\pm$ bosons decaying to leptons can be expressed as
\begin{equation}
D^{\lambda}_{\lambda'}=l_{\lambda}l_{\lambda'},\qquad
        \bar D^{\bar\lambda}_{\bar\lambda'}= \bar l_{\bar\lambda}\bar l_{\bar\lambda'}.
\end{equation}
From the above, we derive the angular distribution. Since we focus solely on the azimuthal angle distribution in our study, we encapsulate the polar angle dependence in coefficients, resulting in an angular distribution of
\begin{equation}\label{eq:ProductionDecayDensitymatrix}
\begin{aligned}
        \frac{d\sigma}{d\Omega^+ d\Omega^-}&=\frac{1}{N}\mathscr{P}^{\lambda\bar\lambda}_{\lambda'\bar\lambda'}D^{\lambda}_{\lambda'}\bar D^{\bar\lambda}_{\bar\lambda'}\\
        &=\frac{1}{N}\Big[ \mathcal{A}+\Big\{ 2\operatorname{Re}[ \mathcal{B} \cos (\phi)+\bar{\mathcal{B}} \cos(\bar{\phi})+\mathcal{C} \cos(2 \phi)+\bar{\mathcal{C}} \cos( 2 \bar{\phi}) \\
& +\mathcal{D}_{ \pm} \cos (\phi \pm \bar{\phi})+\mathcal{E}_{ \pm} \cos (2 \phi \pm \bar{\phi})+\bar{\mathcal{E}}_{ \pm} \cos (\phi \pm 2 \bar{\phi}) \\
& \left.\left.+\mathcal{G}_{ \pm} \cos (2 \phi \pm 2 \bar{\phi})\right]+(\operatorname{Re} \rightarrow \mathrm{Im}) + (\cos \rightarrow \sin )\right\}\Big],
\end{aligned}
\end{equation}
Here, $N=9/(64\pi^2)$ is the normalization factor, and the coefficients ($\mathcal{A}, \mathcal{B}, \bar{\mathcal{B}},...$) are functions of the polar angles of the leptons.  The relevant coefficients are
\begin{equation}
\begin{aligned}
    &\mathcal{A}=\mathscr{P}^{\lambda\bar\lambda}_{\lambda\bar\lambda}d^2_{\lambda}d^2_{\bar \lambda},\\
    &\mathcal{D}_{-}=d_0 \bar d_0 \Big[\mathscr{P}_{++}^{00}d_{-}\bar d_{-} + \mathscr{P}_{+0 }^{0-}d_-\bar d_+ + \mathscr{P}_{0+ }^{-0}d_+\bar d_- +\mathscr{P}_{00 }^{--}d_{+}\bar d_{+}\Big], \\ 
     &\mathcal{G}_-=d_+d_-\bar{d}_+\bar{d}_-\mathscr{P}^{--}_{++}.
\end{aligned}
\end{equation}
If we denote the relative decay azimuthal angle as $\Delta\phi= \phi-\bar\phi$, then $\mathcal{D}_{-}$ is associated with $\Delta\phi$ and $\mathcal{G}_-$ is associated with $2\Delta\phi$.  After integrating over $\phi+\bar\phi$ and the polar angles in the $\mathcal{D}_{-}$ and $\mathcal{G}_{-}$ coefficients, we obtain the distribution of the relative azimuthal angle $\Delta\phi$,
\begin{equation}
\begin{aligned}
    \frac{1}{\sigma}\frac{d\sigma}{d\Delta\phi} = \frac{1}{2\pi} +  
    A \cos(\Delta\phi) +
    \tilde{A} \sin(\Delta\phi) +B \cos(2\Delta\phi) +
    \tilde{B} \sin(2\Delta\phi),
\end{aligned}
\end{equation}
where the coefficients are  
\begin{align}
    A&=\frac{9\pi}{128}\operatorname{Re}\left(\mathscr{P}^{00}_{++}  +\mathscr{P}^{0-}_{+0}+\mathscr{P}^{-0}_{0+}+\mathscr{P}_{00}^{--}\right)  , \\
    \tilde{A}&=\frac{9\pi}{128}\operatorname{Im}\left(\mathscr{P}^{00}_{++}  +\mathscr{P}^{0-}_{+0}+\mathscr{P}^{-0}_{0+}+\mathscr{P}_{00}^{--}\right), \\
   B&=\frac{1}{4\pi}\operatorname{Re}\mathscr{P}_{++}^{--}  , \\
    \tilde{B}&=\frac{1}{4\pi}\operatorname{Im}\mathscr{P}_{++}^{--} .    
\label{eq:azimuthal_coefficients}
\end{align}
The coefficients of the leading terms $\sin(\Delta\phi)$ and $\cos(\Delta\phi)$ can be written in the form of spin correlation coefficients,  
\begin{align}
    A&=-\frac{9\pi}{64}\left(\tr(C_{ij})-C_{33}\right)= -\frac{9\pi}{64}\left(C_{ii}-C_{ij}z_iz_j\right) , \\
    \tilde{A}&=-\frac{9\pi}{64}\left(C_{12}-C_{21}\right)= -\frac{9\pi}{64}\epsilon_{ijk}C_{ij}z_k. 
\end{align}
From the specific form of the coefficients above, we can easily see that these coefficients depend on the choice of the $\vec{z}$ direction. The coefficients of the cosine terms are related to the symmetric part of the spin correlation matrix, while the coefficients of the sine terms are related to the antisymmetric part of the spin correlation matrix. 

\bibliographystyle{apsrev4-1}
\bibliography{ref}
\end{document}